\documentclass[reprint,superscriptaddress,
amsmath,
amssymb,
aps,
prb,
floatfix,
]{revtex4-2}

\usepackage{lipsum}

\usepackage[justification=Justified, format=plain, font=small]{caption}
\usepackage[justification=Justified, format=plain, font=small]{subcaption}
\usepackage{braket}
\usepackage{color}
\usepackage{ulem}
\usepackage{appendix}

\usepackage{graphicx}
\usepackage{dcolumn}
\usepackage{bm}
\usepackage[colorlinks=true, linkcolor=blue, urlcolor=blue,citecolor=blue]{hyperref}
\usepackage{color}
\usepackage{here}
\usepackage{txfontsb}
\usepackage{mathcomp}
\usepackage{dsfont}

\usepackage{xcolor}
\colorlet{myred}{red!40}
\colorlet{myblue}{blue!40}
\colorlet{mycyan}{cyan!40}
\colorlet{mygreen}{green!40}

\newcommand{\red}[1]{{#1}}

\newcommand{\Stiefel}{\text{St}(n,m)}

\definecolor{mygray}{gray}{0.7}

\def\tum{{Technical University of Munich, TUM School of Natural Sciences, Physics Department, 85748 Garching, Germany}}
\def\mcqst{{Munich Center for Quantum Science and Technology (MCQST), Schellingstr. 4, 80799 M{\"u}nchen, Germany}}
\def\utokyo{Department of Applied Physics, the University of Tokyo, 7-3-1 Hongo, Bunkyo-ku, Tokyo 113-8656, Japan}
\newcommand{\TUM}{\affiliation{\tum}}
\newcommand{\MCQST}{\affiliation{\mcqst}}
\newcommand{\UTOKYO}{\affiliation{\utokyo}}

\usepackage{mathtools}
\mathtoolsset{showonlyrefs}

\DeclareMathAlphabet{\mathbbb}{U}{bbold}{m}{n}
\DeclareMathOperator{\Tr}{Tr}
\DeclareMathOperator{\qf}{qf}
\DeclareMathOperator{\argmax}{argmax}

\newcommand{\id}{\mathbbb{1}}

\begin{document}

\title{Holographic Representation of One-Dimensional Many-Body Quantum States via Isometric Tensor Networks}
\author{Kaito Kobayashi}\thanks{These authors contributed equally to this work.} \UTOKYO
\author{Benjamin Sappler}\thanks{These authors contributed equally to this work.} \TUM \MCQST
\author{Frank Pollmann} \TUM \MCQST
\date{\today}

\begin{abstract}
\red{Tensor network methods, most prominently matrix product states (MPS), have become fundamental tools in modern quantum many-body physics.}
\red{While MPS and extensions like the multiscale entanglement renormalization ansatz (MERA) and tree tensor networks (TTN) efficiently capture area-law entanglement and its logarithmic violations, respectively, they inherently struggle to represent highly entangled wavefunctions.} 
\red{Specifically, although central to many frontier problems, reaching the volume-law regime typically demands exponential resources within these conventional frameworks.} 
\red{Motivated by this challenge, we propose holographic isometric tensor network states (holographic isoTNS) that simulate quantum lattice models in \(D\) spatial dimensions via \((D+1)\)-dimensional networks of tensors.} 
\red{The additional dimension substantially enlarges the representational manifold, while isometric constraints on each tensor ensure efficient contractibility.} 
\red{Using one-dimensional systems as testbeds, we analyze the properties of holographic isoTNS.} 
First, we show that randomly initialized holographic isoTNS typically display volume-law entanglement \red{at fixed} bond dimension, surpassing the representational limits of MPS\red{, MERA, TTN,} and related ansätze.
Second, through analytic constructions and variational optimization, we demonstrate that holographic isoTNS can faithfully \red{describe a broad class of highly entangled yet low-complexity states.} 
\red{In particular, the ansatz can represent} arbitrary fermionic Gaussian states, Clifford states, \red{extensions of rainbow states,} and certain short-time-evolved states under local evolution\red{, spanning both classically simulable regimes and those beyond the efficient reach of conventional methods.} 
Third, to exploit this expressivity in \red{broader contexts}, we implement a time-evolving block decimation (TEBD) algorithm on holographic isoTNS.
While the method remains efficient and scalable, \red{error accumulation over TEBD sweeps suggests the need for further algorithmic improvement.}
Overall, holographic isoTNS broaden the \red{scope} of tensor-network methods, opening new avenues to study physics in the volume-law regime.
\end{abstract}

\maketitle

\section{Introduction}
Quantum many-body physics lies at the heart of modern science, yielding emergent phenomena ranging from high-temperature superconductivity to exotic topological phases~\cite{Anderson:MoreIsDifferent,Lee:RevModPhys:2006,Hasan:RevModPhys:2010}. 
Yet the very source of that diversity---the exponential dimensionality of the many-body Hilbert space---makes a full description of quantum states notoriously difficult. 
Among the strategies devised to tame this complexity, tensor network representations have proven exceptionally fruitful~\cite{Cirac:RevModPhys:2021}. 
In one dimension (1D) in particular, matrix product states (MPS) combined with density matrix renormalization group (DMRG) algorithms have become the gold standard for ground-state and low-energy calculations~\cite{White:PRL:1992,White:PRB:1993,Schollwock:RevModPhys:2005,Verstraete:PRB:2006,Hastings:JStatMech:2007,Verstraete:AdvPhys:2008, Ulrich:AnnPhys:2011,Orus:AnnPhys:2014}. 
This efficiency, however, comes at the cost of restricting the accessible manifold of states \red{both in entanglement and complexity~\cite{complexity}.} 
\red{In particular, MPS provides an efficient description for} weakly entangled (``area-law") and low-complexity states, \red{and} the multiscale entanglement renormalization ansatz (MERA) and tree tensor networks (TTN) broaden this reach \red{by capturing} the critical (logarithmic) scaling of entanglement entropy~\cite{Shi:PRA:2006,Vidal:PRL:2007,Vidal:PRL:2008,Murg:PRB:2010}.
Nevertheless, these approaches become prohibitively costly \red{for highly entangled (``volume-law") states, requiring a number of parameters that scales exponentially with the system size.} 
\red{Within the volume-law regime}, highly entangled, high-complexity states \red{are often of limited physical relevance,} as they arise only after exponentially long evolution times~\cite{Poulin:PRL:2011,Brandao:PRXQ:2021}\red{.} 
\red{In contrast,} highly entangled yet low-complexity states \red{are} of \red{much} greater physical interest, \red{such as those obtained after short-time evolution or those with nontrivial structural constraints}~\cite{Sheng-Hsuan:PRXQ:2021}. 

A natural way to \red{access such conventionally unreachable regimes} is to enhance the network variational manifold by enlarging the network configuration; however, this quickly drives up the computational cost. 
Useful guidance comes from isometric tensor network states (isoTNS), which efficiently represent quantum states \red{in two dimensions (2D) and higher}~\cite{Zaletel:PRL:2020, Haghshenas:PRB:2020, Hyatt:arXiv:2019,Benjamin:arXiv:2025}. 
\red{This efficiency is nontrivial because, in these higher dimensions}, the presence of internal loops makes exact network contraction exponentially costly~\cite{Verstraete:2004,Verstraete:PRL;2006,Schuch:PRL:2007,Itai:SIAM:2010}. 
IsoTNS mitigates this scaling by imposing isometric constraints on the tensors, such that many internal tensor contractions reduce to identity operations. 
\red{Crucially}, a 1D tensor network with such isometric conditions is precisely an MPS in isometric form (sometimes also referred to as canonical form~\cite{Vidal:PRL:2003,Vidal:PRL:2004}), which motivated the higher-dimensional extension to isoTNS. 
Here, we bring the isoTNS philosophy back to 1D\red{.} 
We present tensor networks with geometries richer than a simple chain, capable of representing certain highly entangled 1D quantum states\red{, while retaining computational efficiency through isometric constraints}. \par

In this work, we propose a holographic isoTNS ansatz that represents a 1D quantum state on a (1+1)-dimensional lattice of isometric tensors. 
One dimension corresponds to the physical spatial axis; the second, which we view as a ``virtual time" \red{or ``holographic"} axis \red{in analogy with holography}, endows the ansatz with additional expressive power. 
\red{T}he isometric condition imposed on every tensor prevents the exponential blow-up of computational costs that would otherwise accompany this enlarged geometry. 
Importantly, \red{this holographic structure efficiently captures volume-law entanglement using a number of parameters that scales only quadratically with the system size, thereby going beyond conventional ansätze.} 
\red{To further characterize this expressive power, we present} analytical arguments and variational benchmarks\red{.} 
\red{We show} that holographic isoTNS efficiently represents a broad class of physically significant wave functions characterized by high entanglement \red{and} low complexity. 
\red{Examples include} fermionic Gaussian states (FGS)~\cite{Surace:SciPostPhysLectNotes:2022}, Clifford states~\cite{gottesman_theory_1998}, \red{extensions of rainbow states~\cite{Dasgupta:PRB:1980},} and certain short-time-evolved states\red{.} 
\red{Some of these states are classically tractable and others are not, but holographic isoTNS can represent all of these examples within a single ansatz.} 
Furthermore, we implement \red{a} time-evolving block decimation (TEBD) algorithm for real-time evolution \red{on our ansatz}~\cite{Vidal:PRL:2004,Zaletel:PRL:2020}. 
\red{However, we find that accumulated} errors during TEBD evolution \red{currently} limit the \red{use} of the ansatz's high expressive power\red{,} \red{leaving algorithmic refinement for future work.} 
\red{T}he holographic isoTNS ansatz opens a pathway toward exploration of highly entangled quantum states, advancing a frontier of quantum many-body physics beyond conventional ansätze. \par
The remainder of this paper is organized as follows. 
In Sec.~\ref{sec2}, we briefly review MPS and isoTNS in 2D, after which we formally introduce holographic isoTNS and explain their relationship to quantum circuits. 
In Sec.~\ref{sec3}, we closely examine the representational power of holographic isoTNS based on analytical and variational approaches. 
In Sec.~\ref{sec4}, we implement the TEBD algorithm on holographic isoTNS and benchmark this method against exact time evolution. 
Finally, Sec.~\ref{sec5} is devoted to discussions and conclusions.

\section{Holographic isoTNS}\label{sec2}

Let us begin with a general representation of a quantum many-body state. A pure state with \(L\) local degrees of freedom\red{,} such as lattice sites or particles\red{,} can be expressed as
\begin{equation}
    |\Psi\rangle = \sum_{j_1=1}^{d_1}\sum_{j_2=1}^{d_2}\dots\sum_{j_L=1}^{d_L} \psi^{j_1, j_2,\dots, j_L} |j_1\rangle \otimes |j_2\rangle\otimes\dots\otimes|j_L\rangle, \label{psi}
\end{equation}
where \(\{\ket{j_n}\}\) \((j_n=1,\dots, d_n)\) spans the \(d_n\)-dimensional local Hilbert space \(\mathcal{H}_n\) of site \(n\). 
The set \(\{|j_1\rangle\otimes\cdots\otimes|j_L\rangle\}\) constitutes an orthonormal basis of the full Hilbert space \(\mathcal{H}=\bigotimes_{n=1}^{L}\mathcal{H}_n\), and the complex coefficients \(\psi^{j_1, j_2,\dots, j_L}\) fully specify the many-body state.
In the following, we set \(d_n=d\) for simplicity. 
Since the dimension of \(\mathcal{H}\) grows exponentially as \(d^L\), exact treatments quickly become infeasible as \(L\) increases. 
Tensor network representations address this hurdle by truncating the state in a controlled way, decomposing \(\psi^{j_1, j_2,\dots, j_L}\) into a network of low rank tensors.

\subsection{Brief review of isometric MPS}

\begin{figure}[t]
    \centering
    \includegraphics[width=\hsize]{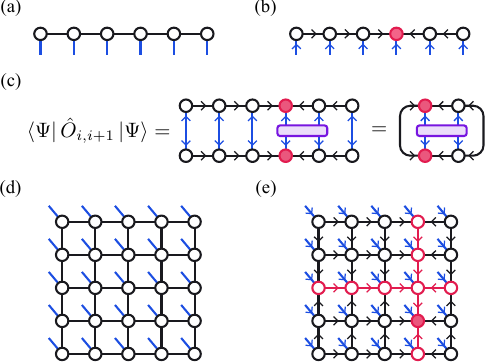}
    \caption{
    Graphical notation for tensor networks. 
    Each circle is a tensor, with blue bonds for physical legs and black bonds for virtual legs connecting the tensors. 
    Arrows indicate isometries. 
    (a) Generic MPS. 
    (b) MPS in isometric form, where all tensors are isometries directed toward the orthogonality center (filled red). 
    (c) Contraction scheme for the local two-site expectation value \(\langle\Psi|\hat{O}_{i,i+1}|\Psi\rangle\). 
    Due to the isometric form, most of the network contracts to identity, leaving a non-trivial contraction of only four tensors. 
    (d) PEPS representing a 2D quantum state on the square lattice. 
    (e) IsoTNS for the same 2D system, where the orthogonality surface is drawn in red and the orthogonality center tensor is filled red. 
    }
    \label{fig1}
\end{figure}

MPS are a cornerstone of tensor network theory~\cite{Verstraete:AdvPhys:2008, Ulrich:AnnPhys:2011,Orus:AnnPhys:2014,cirac_matrix_2021, banuls_tensor_2023, orus_tensor_2019, bridgeman_hand-waving_2017}. 
An MPS is defined by recasting the wave function coefficients as
\begin{equation}
    \psi^{j_1, j_2,\dots, j_L} = \sum_{\alpha_1=1}^{\chi_1}\sum_{\alpha_2=1}^{\chi_2}\dots\sum_{\alpha_{L-1}=1}^{\chi_{L-1}} A^{[1],j_1}_{1,\alpha_{1}}A^{[2],j_2}_{\alpha_{1},\alpha_{2}}\dots A^{[L],j_L}_{\alpha_{L-1},1}, \label{MPS}
\end{equation}
where each rank-\(3\) tensor \(A^{[n],j_n}_{\alpha_{n-1},\alpha_{n}}\) has a physical index \(j_n\) that specifies a basis state on site \(n\) and two virtual indices \(\alpha_{n-1}\) and \(\alpha_n\) whose dimensions (i.e., bond dimensions) are \(\chi_{n-1}\) and \(\chi_n\), respectively (with \(\chi_{0}=\chi_{L}=1\)). 
\red{In principle, any quantum state can be written as an MPS, but f}or generic \red{cases,} the required bond dimension \(\chi_n\) grows exponentially with the system size. 
Computations therefore rely on an approximate form in which every bond is truncated to a fixed maximum dimension \(\chi\). 
This dramatically reduces the number of parameters from exponential to linear in \(L\). 
Typically, truncation is carried out via singular value decompositions (SVDs) that discard the smallest Schmidt coefficients across each bond. 
The price paid is an upper bound on the entanglement: the entanglement entropy is upper bounded by \(S \le \log \chi\) for any bipartition, where \(S =-\mathrm{Tr}\rho_A\log\rho_A\) is the von Neumann entropy of the reduced density matrix \(\rho_A\). 
Therefore, while MPS faithfully describe low entangled states, more general states with volume-law entanglement are outside the efficient reach of MPS \red{at a fixed bond dimension}. 
Nevertheless, because ground states of gapped, local Hamiltonians obey area-law scaling~\cite{Verstraete:PRB:2006,Hastings:JStatMech:2007}, MPS have proven to be a powerful tool for exploring the low-energy physics of quantum many-body systems. 

To obtain stable and fast MPS algorithms, it is advantageous to impose isometric constraints on the tensors~\cite{Vidal:PRL:2003,Vidal:PRL:2004}. 
A matrix \(W\in \mathbb{C}^{n\times m}\) with \(n\ge m\) is an isometry if
\begin{equation}
    W^\dagger W = \mathds{1}_{m}, \quad W W^\dagger = \mathbb{P}_{n}, \label{IsometryCondition}
\end{equation}
where \(\mathds{1}_{m}\) is the identity on the smaller subspace and \(\mathbb{P}_n\) is a projector from the larger subspace into the smaller one (\(\mathbb{P}_n^2=\mathbb{P}_n\)). 
When \(n=m\), isometries correspond to unitaries. 
The definition extends naturally to higher rank tensors by grouping indices. 
Throughout this work we employ the usual tensor network diagram notation: tensors appear as nodes (circles) and their indices as legs; contracting two legs denotes a summation over the shared virtual index. 
As an example, we draw the MPS \red{given by} Eq.~\eqref{MPS} \red{using} tensor diagram notation in Fig.~\ref{fig1}a. 
Additionally, to visually represent isometries, we decorate the legs with arrows, pointing from the larger subspace dimension towards the smaller one. 
This provides a compact diagrammatic representation of the condition in Eq.~\eqref{IsometryCondition}:
\begin{equation}
    \vcenter{\hbox{\includegraphics[scale=1]{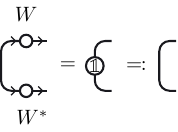}}}, \quad\quad \vcenter{\hbox{\includegraphics[scale=1]{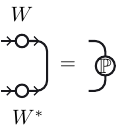}}}.
\end{equation}
\par
By consecutively applying QR decompositions or SVDs, any MPS can be brought into its \textit{isometric form} in Fig.~\ref{fig1}b: a single site is selected as the orthogonality center, while every other tensor is taken to be isometric with bond arrows all directed towards the orthogonality center~\red{\cite{Ulrich:AnnPhys:2011}}.
It is worth emphasizing that the isometric form of MPS is exact, i.e., every MPS can be brought into isometric form without error. 
Shifting the orthogonality center is purely a gauge choice and can be carried out without introducing error. 
This isometric form is computationally advantageous in both efficiency and numerical stability~\red{\cite{Ulrich:AnnPhys:2011}}.
For example, when computing the norm \(\langle \psi|\psi\rangle\), the isometric tensors cancel pairwise, leaving only a single contraction at the orthogonality center. 
An analogous simplification applies to a local expectation value \(\langle \psi|O_{i,i+1}|\psi\rangle\): one first moves the orthogonality center to site \(i\) or \(i+1\), after which the isometry condition can be used to reduce the computation to a contraction of only four tensors and the gate itself, as shown in Fig.~\ref{fig1}c. 

\subsection{IsoTNS in 2D quantum systems}~\label{sec2-2}

Tensor network representations can be extended naturally to higher dimensions. 
In 2D, MPS is generalized to projected entangled-pair states (PEPS)~\cite{Verstraete:2004}. 
As depicted in Fig.~\ref{fig1}d, a PEPS consists of tensors arranged on a 2D lattice (here a square lattice), where each tensor possesses one physical leg and four virtual legs connecting to its neighbors~\cite{Verstraete:2004,Verstraete:PRL;2006}. 
A key distinction from MPS is that\red{, in general, a} PEPS cannot be brought into an \red{exact} isometric form\red{.}
\red{This is due to the presence of} closed loops in higher-dimensional tensor networks\red{, which obstruct the loop-free gauge fixing available in 1D}. 
\red{In other words, the isometric form is no longer a gauge choice}. 
\red{Note that without such an isometric form}, the cost of an exact contraction in PEPS increases exponentially with system size, meaning that the exact calculation of even local expectation values is generally intractable~\cite{Verstraete:2004,Verstraete:PRL;2006,Schuch:PRL:2007,Itai:SIAM:2010}. 

To overcome this computational barrier, isoTNS was introduced as a structured subclass of PEPS with built-in isometric constraints~\cite{Zaletel:PRL:2020, Haghshenas:PRB:2020, Hyatt:arXiv:2019,Benjamin:arXiv:2025}. 
As shown in Fig.~\ref{fig1}e, the network is constructed from a set of directed isometric tensors: the legs of bulk tensors point toward a designated row and column, which is called the orthogonality surface, and tensors on the orthogonality surface point toward a single tensor, which is called the orthogonality center. 
By direct analogy with MPS, expectation values of operators supported at the orthogonality center can be evaluated very efficiently because the surrounding isometries contract to identities~\cite{Zaletel:PRL:2020, Benjamin:arXiv:2025}. 
For example, computing the expectation value of a two-site operator at the orthogonality center requires the contraction of only five tensors including the operator itself. 
The isometric constraints therefore avoid the exponential cost of generic PEPS, yielding a significant improvement in computational efficiency. 
The trade-off is reduced expressive power, since isoTNS form a strict subset of PEPS. 
Nevertheless, pioneering studies show that 2D isoTNS retain much of the expressive power of generic PEPS while greatly reducing computational \red{scaling}~\cite{Zaletel:PRL:2020, Soejima:PRB:2020, Sheng-Hsuan:PRB:2022}. 
Subsequent extensions of the method cover fermionic systems~\cite{Dai:PRL:2025, Yantao:PRXQ:2025}, thermal ensembles~\cite{Wilhelm:PRB:2023}, infinite-strip geometries~\cite{Wu:PRB:2023}, three-dimensional lattices~\cite{Tepaske:PRR:2021}, and quantum circuits~\cite{Slattery:arXiv:2021,Wei:PRL:2022,Malz:PRXQuantum:2025,Leontica:arXiv:2025}.

\subsection{Holographic isoTNS}~\label{sec2-3}

\begin{figure*}[t]
    \centering
    \includegraphics[width=\hsize]{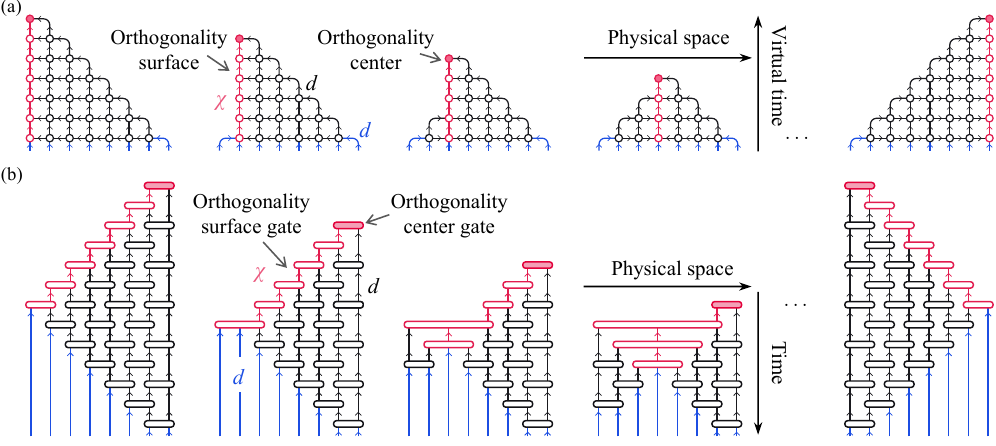}
    \caption{
    (a) Diagrammatic representation of the holographic isoTNS ansatz for $L=8$ with \red{different} positions of the orthogonality surface. 
    The horizontal axis represents physical space, and the vertical axis represents the virtual time domain. 
    Physical legs are shown in blue, and bulk tensors are colored black. 
    The \red{red legs and tensors} represent the orthogonality surface, with the orthogonality \red{center} tensor highlighted as a filled circle. 
    The bond dimension along the orthogonality surface is upper-bounded by \(\chi\), whereas all other bonds are given their full dimension of \(d\). 
    (b) Circuit picture of the ansatz. 
    Each bulk tensor from (a) is mapped to a two-qudit unitary (black) acting on adjacent sites.  
    The isometric tensors are represented by \red{red} unitaries, which have multiple legs and act on fixed ancilla qudits [see Eq.~\eqref{isometrygate}]; for clarity, we omit the ancilla qudits from this visualization, and only show one leg (corresponding to the case with $\chi=2$).
    The execution time of the circuit flows in the opposite direction of the isometric arrows. 
    }
    \label{fig2}
\end{figure*}

We here introduce the holographic isoTNS ansatz, which utilizes a (1+1)-dimensional array of isometrically constrained tensors to represent 1D quantum states. 
The word ``holographic" highlights this dimensional difference \red{by analogy with holography}, distinguishing it from conventional isoTNS where the dimensionality of the tensor network directly matches that of the quantum system (e.g., a 2D isoTNS for a 2D quantum system). 
We note that related holographic connections have been proposed in broader settings, for example between quantum channels and quantum circuits~\cite{Osborne:PRL:2010, Gopalakrishnan:JPhysA:2025, lu_holographic_2025}.

Figure~\ref{fig2}a illustrates the typical structures of holographic isoTNS. 
The horizontal axis represents the spatial dimension of the 1D system, while the vertical axis is interpreted as a virtual time dimension. Let us first focus on the bottom row tensors, which feature physical legs representing a 1D quantum system of length \(L\). 
Each tensor in this layer \red{is} denoted \red{as} \(T_{l,r,a}^{j}\) \red{and} has one physical leg (index \(j\)) and three virtual legs that connect to adjacent tensors on the left (\(l\)), right (\(r\)), and above (\(a\)). 
In the bulk of the network (i.e., higher rows), tensors do not have physical legs\red{.} 
\red{D}enoted \red{as} \(T_{l,r,b,a}\), \red{these tensors} possess four virtual legs connecting to neighbors on the left (\(l\)), right (\(r\)), below (\(b\)), and above (\(a\)). 
We adopt the convention of removing any tensor with only two nontrivial bonds (one incoming, one outgoing), as they can be absorbed into one of their neighboring tensors. 

A central feature of this network \red{is} the \red{isometric constraint}\red{, which assigns a specific isometry direction to each leg of tensors \(T_{l,r,a}^{j}\) and \(T_{l,r,b,a}\).} 
\red{To organize these directions for efficient network contractions (explained later), we define the orthogonality surface as a vertical column of tensors at a fixed site (the red column in Fig.~\ref{fig2}a).} 
\red{For bulk tensors, meaning those outside the orthogonality surface, we choose all horizontal bonds to be isometric toward the orthogonality surface, and all vertical bonds to be isometric upward.} 
Within the orthogonality surface, \red{we further designate} a single tensor as the orthogonality center. 
\red{All other tensors on this surface are vertically} isometrized toward the orthogonality center. 

\red{With these conventions in place, we now describe the network construction.} 
\red{Consider a bottom tensor located at either edge (but not on the orthogonality surface).} 
\red{This tensor receives two physical legs, each carrying \(d\) degrees of freedom, and equally redistributes them through horizontal and vertical legs of dimension \(d\).} 
\red{Neighboring tensors similarly have two incoming and two outgoing legs, each with dimension \(d\).} 
\red{This process continues until reaching the orthogonality surface.} 
\red{By design, these bulk tensors propagate} information across the spatial and virtual time domains without any approximations (truncations) \red{as they uniformly maintain the exact dimension \(d\).} 
\red{At the orthogonality surface,} information \red{cascading} through the \red{bulk accumulates}. 
\red{Achieving an exact representation of this accumulation necessitates an exponential growth} in the dimensions of the vertical bonds along the orthogonality surface. 
\red{To manage this, we truncate these vertical bonds} to a maximum bond dimension \(\chi\) \red{using standard MPS algorithms.} 
\red{This completes the construction of holographic isoTNS shown in Fig.~\ref{fig2}a.} 
\red{For implementation details, we also refer to our source code~\cite{kobayashi_2025_17907383}.}\par
Importantly, the isometric constraints ensure efficient tensor operations for local updates at the orthogonality surface. 
\red{Similar to contractions for MPS in an isometric form (Fig.~\ref{fig1}c), contracting any isometrized column with its complex conjugate yields the identity.} 
\red{This property reduces the full network evaluation to a local contraction involving only the columns affected by the operator.} 
A key example is the calculation of a local two-site expectation value, which can be dramatically simplified as
\begin{equation}
\vcenter{\hbox{\includegraphics[scale=1]{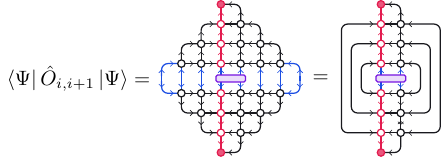}}},
\end{equation}
where the operator \(\hat{O}_{i,i+1}\) is represented as a purple gate. 
The computational cost of this contraction scales as ${\mathcal{O}(L\chi^3)}$ \red{(linearly with \(L\))}.
The same simplification applies when evaluating the wave function norm \(\langle\Psi|\Psi\rangle\), where the isometric constraints cancel every tensor except the orthogonality center. 
Normalizing the state therefore amounts to requiring that the orthogonality center be a tensor of norm one. \par
\red{This computational efficiency, however, cannot be leveraged directly for operators acting away from the orthogonality surface.} 
\red{In such cases,} the orthogonality surface must first be shifted to the operator's location. 
The systematic procedure for this shift is detailed in Sec.~\ref{sec4-1} and Appendix~\ref{app:MM}. 
\red{Crucially, the position of the orthogonality surface is not merely a gauge choice.} 
\red{While one can freely shift the orthogonality \textit{center} along the orthogonality \textit{surface} (just as in standard MPS), the position of the orthogonality \textit{surface} itself fundamentally alters the network's structure, as illustrated in Fig.~\ref{fig2}a.} 
\red{As a result, this placement does affect the manifold of states representable by the ansatz.} 
\red{Despite this, because this position is an auxiliary degree of freedom to facilitate computations, it should have no physical meaning.} 
\red{It is therefore natural to conjecture that, for physically relevant states (i.e., those that do not intrinsically reflect the network's geometric details), the representational power of the ansatz should not strongly depend on where the orthogonality surface is placed.} 
\red{Indeed, we show in Sec.~III that the position of the orthogonality surface does not affect the representability of certain classes of states.} 
\red{Consequently, in what follows we evaluate the ansatz while fixing the orthogonality surface at representative positions.} 
\red{Precisely characterizing how the representable manifold varies with the position of the orthogonality surface remains a largely open question.}

\red{Another important point is} that a holographic isoTNS can be directly mapped onto a quantum circuit, as shown in Fig.~\ref{fig2}b. 
The principle underlying this mapping is that any isometry can be extended to a unitary operator on a larger Hilbert space by incorporating ancillary degrees of freedom~\cite{Schon:PRL:2005,Slattery:arXiv:2021, Wei:PRL:2022}. 
Schematically, this conversion is illustrated as 
\begin{equation}
    \vcenter{\hbox{\includegraphics[scale=1.0]{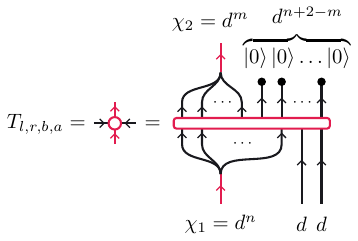}}},\label{isometrygate}
\end{equation}
where the input and output legs with dimensions \(\chi_1=d^n\) and \(\chi_2=d^m\) are resolved into \(n\) and \(m\) qudit wires, respectively. 
By adjoining \(n+2-m\) ancillary qudits, the isometry extends to a unitary gate acting on \(n+2\) qudits. 
The ancillary qudits are subsequently contracted with a reference state, thereby realizing the original isometry. 
For MPS in the isometric form (Fig.~\ref{fig1}b), sweeping this embedding along the chain yields a unitary circuit that effectively prepares the state from ancillas initialized in a reference state. 
By convention, the flow of the circuit execution time is opposite to the orientation of the isometry arrows. 
\red{T}he representational power of quantum states generated by such tensor-network-inspired quantum circuit\red{s} has been a subject of active recent study~\cite{Sheng-Hsuan:PRB:2022,Haghshenas:PRX:2022,Sajant:PRXQ:2023,Malz:PRL:2024}. 

In holographic isoTNS, the isometries on the orthogonality surface are similarly converted to unitary gates according to Eq.~\eqref{isometrygate}. 
Concurrently, the bulk isometric tensors are directly interpreted as two-qudit unitary gates, as they possess two incoming and two outgoing indices of the same bond dimension \(d\). 
The composition of these two types of gates forms the full circuit representation shown in Fig.~\ref{fig2}b, making this ansatz useful in the context of quantum computing. 
As in the tensor network picture, the resulting circuit architecture is determined by the location of the orthogonality surface. 
Placing it at the boundary produces a triangle-shaped circuit, whereas positioning it near the middle introduces intrinsically long-range isometric gates. 
In either case, the circuit first prepares the state by acting with isometry gates on ancilla qudits, followed by a unitary network whose depth scales as ${\mathcal{O}(L)}$.

\red{We close by discussing the connection of the holographic isoTNS to TTN~\cite{Shi:PRA:2006, Murg:PRB:2010} and to MERA~\cite{Vidal:PRL:2007, Vidal:PRL:2008}.} 
\red{Similar to our construction, both TTN and MERA represent $D$-dimensional quantum states by means of a $(D+1)$-dimensional tensor network geometry.} 
\red{The main structural difference lies in how the network height scales with the system size $L$: logarithmically for TTN and MERA, but linearly for holographic isoTNS}.
\red{This distinction reflects their fundamentally different relationships to the renormalization group (RG).} 
\red{Both TTN and MERA possess a well-established connection to RG~\cite{Levin:PRL:2007,Xie:PRB:2012, evenbly_tensor_2015-1, evenbly_tensor_2015}, where the additional network direction is explicitly interpreted as an RG scale.} 
\red{This naturally leads to logarithmic scaling in height and a logarithmic upper bound on entanglement.} 
\red{By contrast, the holographic isoTNS is not derived from a renormalization flow.} 
\red{Its additional dimension does not represent an RG scale, nor is the network structured as a hierarchical coarse-graining transformation.} 
\red{Consequently, its height scales linearly with \(L\), rather than exhibiting the conventional logarithmic structure associated with RG flows (although linear coarse-graining schemes have been explored in certain contexts~\cite{lan_tensor_2019}).}

\section{Representation power of holographic isoTNS}\label{sec3}

\red{In this section, we} investigate the representational power of holographic isoTNS. 
\red{T}hrough analytical constructions and variational calculations we demonstrate that \red{the ansatz} provide\red{s} an efficient representation for several key families of quantum many-body states. 
\red{Hereafter,} we set the local physical dimension \red{to} \(d=2\). 

\subsection{Entanglement volume\red{-}law of typical states}\label{sec3-1}

\begin{figure}[t]
    \centering
    \includegraphics[width=\hsize]{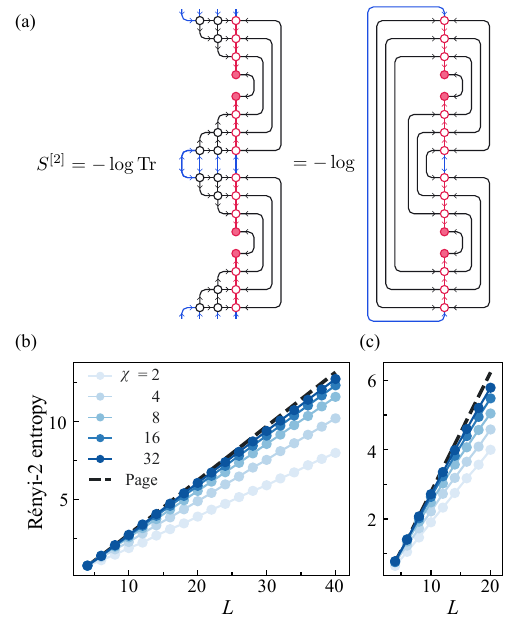}
    \caption{
        (a) Contraction scheme for the calculation of the half-chain second R\'{e}nyi entropy when the orthogonality surface is placed at the center of the system. 
        (b), (c) R\'{e}nyi-2 entanglement entropy \(S^{[2]}\) as a function of the system size \(L\). 
        The blue lines represent results for holographic isoTNS with varying the bond dimension \(\chi\), and the black dashed line plots the Page value. 
        The orthogonality surface is placed at \(L/2\) for (b) and at \(1\) for (c). 
        The data is averaged over \(128\) random realizations. 
    }
  \label{fig3}
\end{figure}

First, we examine the entanglement properties of typical holographic isoTNS. 
We generate an ensemble of random holographic isoTNS by initializing every tensor with random coefficients drawn from the Haar distribution, subject to the required isometric constraints. 
For each realization, we evaluate the second R\'{e}nyi entropy \(S^{[2]} = -\log\mathrm{Tr}\rho_A^2\) for a half-chain bipartition, where \(\rho_A\) is the reduced density matrix of one half of the system~\cite{Lennert:JMathPhys:2013}. 
\red{Placing} the orthogonality surface at the chain midpoint \red{enables the efficient calculation of \(S^{[2]}\), because the isometry conditions simplify} the required tensor contraction \red{as} shown in Fig.~\ref{fig3}a. 

Figure~\ref{fig3}b shows the ensemble-averaged second R\'{e}nyi entropy for this choice of the orthogonality surface, varying the system size \(L\) and bond dimension \(\chi\). 
Remarkably, the entanglement grows linearly with \(L\) in all cases. 
This demonstrates that typical (random) states of holographic isoTNS exhibit volume-law entanglement. 
Increasing the bond dimension \(\chi\) results in a steeper linear growth, which systematically approaches the reference entropy for random pure states \(S^{[2]} \sim (L/2 -1) \log 2\), known as the Page value~\cite{Lubkin:JMathPhys:1978,Page:PRL:1993,Foong:PRL:1994}. 

When the orthogonality surface is placed away from the midpoint (\(\lfloor L/2\rfloor\) or \(\lfloor L/2+1\rfloor\)), the shortcut contraction of Fig.~\ref{fig3}a is no longer applicable. 
For small systems, however, we can still obtain the second R\'{e}nyi entropy by fully contracting the network into a \(2^L\)-dimensional state vector and then computing \(S^{[2]}\) directly. 
Figure \ref{fig3}c plots the resulting \(S^{[2]}\) for randomly initialized holographic isoTNS with the orthogonality surface fixed at site \(1\) (leftmost). 
The entanglement exhibits clear volume-law scaling, similar to the results in Fig.~\ref{fig3}b. 
We confirm that this volume-law behavior persists for other choices of the orthogonality surface location. \par
This capacity to represent extensive entanglement is an intrinsic feature of the (1+1)-dimensional architecture sketched in Fig.~\ref{fig2}a. 
For a half-chain bipartition, the number of horizontal bonds cut along the spatial dimension grows linearly with the system size \(L\). 
Crucially, these spatial bonds are never truncated; all truncations occur only on the orthogonality surface, where bonds run vertically in the virtual time direction. 
Because each untruncated spatial bond can carry entanglement, the total entanglement entropy can scale linearly with \(L\)---a volume\red{-}law in 1D. 
This behavior contrasts sharply with MPS, where the two halves of the system are connected by a single spatial bond. 
Truncating that lone bond rigidly bounds the entanglement entropy, precluding any volume\red{-}law growth. 
Therefore, the ability to capture volume-law entanglement is one of the most significant characteristics of holographic isoTNS.

\subsection{Exactly representable classes of states}\label{sec3-2}

Building on the discussion of entanglement above, we now analytically and variationally demonstrate representability of holographic isoTNS for concrete families of states. 

{\it MPS.}
---We begin with two canonical examples: the Greenberger–Horne–Zeilinger (GHZ) state \cite{Greenberger:AmJPhys:1990} and the W state \cite{W:PRA:2000}. 
Defined as \(\ket{\mathrm{GHZ}}=\frac{1}{\sqrt{2}} (\ket{00\dots 0} + \ket{11\dots 1})\) and \(\ket{\mathrm{W}}=\frac{1}{\sqrt{L}}(\ket{10\dots 0}+\ket{01\dots 0} +\dots+\ket{00\dots 1})\), respectively, both states have Schmidt rank \(2\) across any nontrivial bipartition. 
This allows them to be represented exactly by MPS with bond dimension \(\chi=2\). 
Notably, the holographic isoTNS architecture naturally accommodates such states. 
By setting all tensors in rows above the physical layer to identity operators, the whole network effectively reduces to 1D MPS in the bottom row with bond dimension \(\chi=2\). 
Consequently, holographic isoTNS can exactly represent the GHZ and W states\red{,} and any state with Schmidt rank \(2\)\red{,} irrespective of the location of the orthogonality surface. 

By rotating our viewpoint by \(90\tcdegree\), we further find that any MPS can be embedded in \red{a} holographic isoTNS. 
Specifically, when the orthogonality surface is placed at the boundary, the target MPS tensors can be encoded vertically on the orthogonality surface, while all bulk tensors are set to the identity. 
Because nontrivial bulk tensors enlarge the variational space, this construction shows explicitly that holographic isoTNS with this surface position are strictly more expressive than MPS at the same bond dimension. 
When the orthogonality surface is positioned elsewhere, some of its tensors receive virtual indices from both horizontal directions. 
Geometrically, this is equivalent to folding an MPS at an intermediate point and merging the two overlapping virtual bonds of dimension \(\chi\) into a single bond of dimension \(\chi^2\). 
Therefore, for arbitrary placement of the orthogonality surface, holographic isoTNS with bond dimension \(\chi^2\) can exactly encode any MPS with bond dimension \(\chi\). 

{\it \red{Extended r}ainbow states.}
---Next, beyond the area-law regime, we present an explicit construction for states with volume-law entanglement. 
The starting point is the circuit representation in Fig.~\ref{fig2}b with the orthogonality surface placed at the leftmost site. 
By configuring a subset of the bulk gates as SWAP gates and the remainder as identity gates, we can implement an arbitrary qubit permutation \([2,3,\dots,L-1,L] \mapsto [\sigma(2), \sigma(3), \cdots, \sigma(L-1), \sigma(L)]\). 
Consequently, the orthogonality surface becomes equivalent to MPS whose physical indices are ordered according to the permutation \((1, \sigma(2), \sigma(3), \cdots, \sigma(L-1), \sigma(L))\). 
For instance, this construction can generate states that are tensor products of entangled two-qubit pairs:
\begin{equation}
    |\Psi\rangle = |\psi_2\rangle_{1,\sigma^{-1}(2)}\otimes |\psi_2\rangle_{\sigma^{-1}(3),\sigma^{-1}(4)}\otimes\cdots|\psi_2\rangle_{\sigma^{-1}(L-1),\sigma^{-1}(L)},\label{twoqubit}
\end{equation}
where \(\sigma^{-1}\) is the inverse permutation and \(L\) is assumed even for clarity. 
\(|\psi_{2}\rangle_{i,j}\) denotes an entangled two-qubit state on qubits \(i\) and \(j\); for bond dimension \(\chi \ge 4\), each \(|\psi_2\rangle_{i, j}\) can be chosen arbitrarily.
Because the permutation \(\sigma\) can entangle arbitrarily distant pairs of qubits, the resulting state \(|\Psi\rangle\) generally exhibits volume-law entanglement. 
A particularly illuminating choice is the \textit{rainbow} permutation: \(\sigma(2m-1) = m\) and \(\sigma(2m)=L-m+1\) for \(1\leq m\leq L/2\). 
This permutation pairs qubits symmetrically across the center of the chain. 
If each pair \(|\psi_{2}\rangle_{i,j}\) is prepared in a maximally entangled singlet state, \(|\Psi\rangle\) \red{reduces to the usual} rainbow state, whose half-chain entanglement saturates the maximal value of \(L/2\log 2\). 
Note that this rainbow state appears as the ground state of the Heisenberg Hamiltonian with inhomogeneous couplings~\cite{Dasgupta:PRB:1980,Vitagliano:NewJPhys:2010,Ramirez:JStatMech:2014}.

{\it Variational optimization.}
---To assess the representational power of holographic isoTNS beyond analytic arguments, we here introduce a variational protocol. 
Given a reference wavefunction \(\ket{\Psi_\text{ref}}\), we optimize all tensors in the ansatz to minimize the distance \(||\ket{\Psi}-\ket{\Psi_\text{ref}}||^2\) using Riemannian variational optimization; see Appendix~\ref{app:riemannian_optimization} for details. 
If the optimizer converges to the configuration that reproduces the target state within the stringent error tolerance \(||\ket{\Psi}-\ket{\Psi_\text{ref}}||^2<10^{-14}\), we take this as positive evidence of representability of the ansatz. 
Conversely, failure to converge may suggest lack of representability, although this cannot be concluded definitively because the optimizer may become trapped in local minima. 
To minimize this possibility, we repeat each optimization with up to 1000 distinct random initializations and take the best result. 
When evaluating the representational power for a given class of states, exhaustive testing is infeasible. 
Instead, we draw 20 reference states at random from the target class and run the above procedure for each reference state. 
If the optimization finds a valid representation for every sampled instance, we regard the class as strongly supported. 
Note that to perform a full variational optimization, the holographic isoTNS needs to be contracted to a full state vector, e.g.,
\begin{equation}
    \vcenter{\hbox{\includegraphics[scale=1.0]{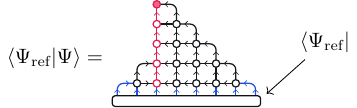}}}\label{contraction_with_state_vector}
\end{equation}
for the computation of the overlap with a reference state $\ket{\red{\Psi}_\text{ref}}$. 
This has a computational cost scaling exponential in system size, limiting the variational approach to a small number of sites. 

{\it FGS.}
---We first target FGS for our variational optimization.
FGS appear as eigenstates of Hamiltonians quadratic in fermionic creation and annihilation operators~\cite{Surace:SciPostPhysLectNotes:2022}. 
After a Jordan-Wigner transformation, the problem moves from fermions to qubits, where each nearest-neighbor free-fermion interaction becomes a two-qubit unitary in a specific form, known as matchgates~\cite{Knill:2001,Valiant:SIAMJComput:2002,Terhal:PRA:2002,Jozsa:2008}.
Consequently, \red{an} FGS can be equivalently characterized as any state prepared by a circuit of matchgates acting on a computational basis state. 
Note that matchgate-only circuits are not universal for quantum computation; moreover, the dynamics of FGSs are efficiently simulable on a classical computer via the covariance matrix formalism~\cite{Jozsa:ProcRSocA:2010,Hebenstreit:PRL:2019}. 
In short, while FGSs may exhibit substantial entanglement, their complexity is limited by their Gaussian structure. 

Based on extensive variational optimization, we find that FGSs are exactly representable by holographic isoTNS with only \(\chi=2\), independent of the location of the orthogonality surface.
In particular, when the orthogonality surface is placed at the boundary, there exists a direct analytic justification via a structural correspondence. 
As established in Ref.~\cite{Raul:arXiv:2025}, any matchgate circuit can be brought into a right-standard form 
\begin{equation}
    \vcenter{\hbox{\includegraphics[scale=1]{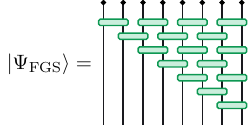}}},
\end{equation}
where the green blocks denote matchgates and the top row of dots represents the computational basis state. 
Strikingly, this circuit layout is structurally identical to a holographic isoTNS with its orthogonality surface at the left-most site (Fig.~\ref{fig2}a). 
This allows for a one-to-one mapping: each matchgate corresponds to a gate in the same position within the holographic isoTNS circuit, and any remaining gates in the holographic isoTNS are taken to be the identity. 
The computational basis state naturally arises within the construction of the isometric gates [Eq.~\eqref{isometrygate}], so the resulting holographic isoTNS exactly reproduces the target FGS with \(\chi=2\).

{\it Clifford states.}
---We next target another prime class of highly entangled yet low complexity states: Clifford states~\cite{gottesman_theory_1998}. 
A Clifford state is generated by the application of a Clifford circuit---composed of Hadamard, phase and CNOT gates---to the computational basis state.
On a linear nearest neighbor architecture, an arbitrary Clifford transformation can be implemented as a Clifford circuit with depth \(\mathcal{O}(7L)\), underscoring the limited complexity of this family~\cite{maslov_shorter_2018, maslov_cnot_2023}. 
Although Clifford states admit efficient classical simulation within the stabilizer formalism, MPS \red{struggle to represent} this class because the states typically exhibit volume-law entanglement.
\red{To evaluate our ansatz against this class, we generate random target Clifford states for system sizes $L=12,13,14$ using the \textbf{qiskit} package~\cite{javadi-abhari_quantum_2024}.} 
We \red{then} perform variational optimization, once with the orthogonality surface at the left border and once with the orthogonality surface in the center of the chain. 
Remarkably, holographic isoTNS successfully represent all target states with bond dimension \(\chi=2\)\red{.} 
\red{This result} strongly \red{supports} the representability \red{of our ansatz} for Clifford states\red{, even in the presence of volume-law entanglement}.

\subsection{Time-evolved states}\label{sec3-4}

An attractive application of holographic isoTNS is time evolution\red{,} $\ket{\Psi(t)}=e^{-i\hat{\mathcal{H}}t}\ket{\Psi_0}$. 
For a generic quantum quench in which the initial state $\ket{\Psi_0}$ is not close to an eigenstate of the Hamiltonian \(\hat{\mathcal{H}}\), the entanglement entropy typically grows linearly in time. 
This rapid entanglement growth necessitates an exponentially increasing bond dimension for accurate MPS representations. 
By contrast, holographic isoTNS can efficiently represent certain volume-law entangled states, suggesting its suitability for time evolution beyond MPS. 

In what follows, we assess the representational power of holographic isoTNS for time-evolved states via \red{full} variational optimization; a direct simulation based on TEBD is presented in Sec.~\ref{sec4-2}. 
\red{To this end, we first \red{simulate} the dynamics via exact diagonalization (ED), yielding a reference state $\ket{\Psi_{\mathrm{ref}}(t)}$ at each time step $t$.} 
\red{Placing the orthogonality surface at the leftmost site, we then variationally optimize the tensors of our ansatz $\ket{\Psi(t)}$ to maximize its overlap with \red{the reference} $\ket{\Psi_{\mathrm{ref}}(t)}$ at every time step.} 
\red{Finally, R\'{e}nyi-2 entropies \(S^{[2]}\) and variational errors \(||\ket{\Psi (t)}-\ket{\Psi_\text{ref} (t)}||^{2}\) are computed by exact contraction of the optimized tensor networks.}

Our first target is the Floquet dynamics of the Kicked Ising Chain (KIC)~\cite{prosen_general_2002, prosen_chaos_2007}
\begin{align*}
    \hat{\mathcal{H}}_\text{KIC} &= \sum_{j=1}^{L}\left[J\hat{\sigma}^{z}_j\hat{\sigma}^{z}_{j+1}+\left(g\hat{\sigma}^x_j+h\hat{\sigma}^z_j\right)\sum_{m=-\infty}^{m=\infty}\delta(t-m)\right] \\
    &\equiv \hat{\mathcal{H}}_\text{I} + \hat{\mathcal{H}}_\text{K}\sum_{m=-\infty}^{m=\infty}\delta(t-m),
\end{align*}
with \(L=14\) and \((J,g,h)=(\pi/4,\pi/4,0.5)\). 
Setting the time between kicks to unity, the time evolution is given by $\ket{\Psi(t+1)}=e^{-i\hat{\mathcal{H}}_K}e^{-i\hat{\mathcal{H}}_I}\ket{\Psi(t)}$. 
We start the evolution from the fully polarized product state $\ket{\Psi_0} = \ket{\uparrow\uparrow\dots\uparrow}$. 
Note that we choose the model parameters such that the KIC cannot be mapped to a fermionic Gaussian model, as would be the case for $h=0$. 
Consequently, \red{the time-evolved state is neither an FGS nor a Clifford state}. 

\begin{figure}[t]
    \centering
    \includegraphics[width=\hsize]{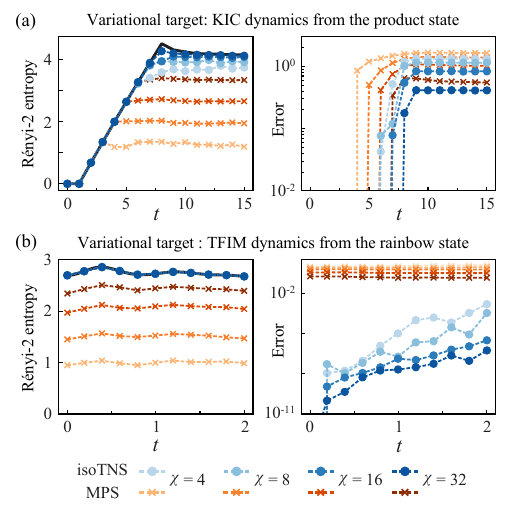}
    \caption{
    Variational optimization of holographic isoTNS (blue) and MPS (red) while changing the bond dimension \(\chi\) (system size \(L=14\)). 
    The left and right panels show, respectively, the R\'enyi-2 entropy \(S^{[2]}\) and the variational error \(||\ket{\Psi}-\ket{\Psi_\text{ref}}||^{2}\) as a function of time. 
    (a) Dynamics under the KIC Hamiltonian at \((J,g,h)=(\pi/4,\pi/4,0.5)\) from the product state $\ket{\Psi_0} = \ket{\uparrow\uparrow\dots\uparrow}$. 
    (b) TFIM dynamics at \((J,g)=(1,1)\) from the highly entangled volume-law rainbow state defined in Eq.~\eqref{twoqubit}.
    For completeness, we note that our results provide only an upper bound on the true variational error, as this computationally hard constrained optimization problem yields only local minima of the cost function.
    }
  \label{fig4}
\end{figure}

Figure~\ref{fig4}a compares the variational performance of MPS and holographic isoTNS in representing time-evolved states of the KIC. 
We plot both the R\'{e}nyi-2 entropy \(S^{[2]}\) and the variational error \(||\ket{\Psi}-\ket{\Psi_\text{ref}}||^{2}\) as function\red{s} of the target evolution time.  \red{Under t}he Floquet driving protocol of the KIC\red{, the system experiences} a rapid, linear growth of entanglement. 
\red{While} the MPS ansatz \red{initially} reproduces the exact dynamics, its accuracy deteriorates abruptly once the entanglement approaches the maximum value sustainable by the chosen bond dimension \(\chi\). 
\red{A} distinct plateau in the calculated \(S^{[2]}\) and a sharp increase in the variational error \red{signal this breakdown}. 
\red{Although i}ncreasing \(\chi\) extends the reliable simulation time\red{, it cannot overcome} this \red{fundamental} entanglement \red{barrier}. 
In contrast, the holographic isoTNS \red{demonstrates a clear advantage for representing quantum dynamics with rapid entanglement growth.} 
\red{It accurately} tracks the entanglement even with small bond dimensions.
\red{While the holographic isoTNS eventually deviates from the exact state at longer times, it delays the onset of this deviation compared to MPS for the same \(\chi\).} 
\red{Note that the captured entanglement \(S^{[2]}\) remains high even after accuracy drops.}

This \red{distinct behavior in the} long-time \red{regime} \red{suggests that the limitations of the two ansätze are fundamentally different}. 
\red{While} the primary bottleneck \red{for MPS} is clearly entanglement growth\red{,} the holographic isoTNS \red{successfully accommodates this large entanglement.} 
\red{Its} \red{eventual deviation therefore} likely reflects a different obstruction.
Because unitary dynamics generically increases not only entanglement but also state complexity~\cite{Brandao:PRXQ:2021,Sheng-Hsuan:PRXQ:2021}, it is natural to attribute the dominant factor limiting the variational power \red{of holographic isoTNS} to the growth of complexity rather than entanglement. 

To distinguish the roles of entanglement and complexity in the variational performance of our ansatz, we investigate a scenario in which the initial state already exhibits volume-law entanglement. 
Unlike evolution from a product state, where both entanglement and complexity increase with time (Fig.~\ref{fig4}a), \red{evolving this pre-entangled state} primarily increases the state complexity while keeping \red{further} entanglement growth moderate. 
\red{Specifically, w}e construct \(\ket{\Psi_0}\) as a product of long-range two-qubit states [Eq.~\eqref{twoqubit}] arranged in the rainbow pairing that connects site \(i\) to site \(L-(i-1)\). 
Each pair is prepared in the state \(\ket{\psi_2}=\sqrt{1/5}\ket{00}+\sqrt{2/5}(\ket{01}+\ket{10})\). 
This construction ensures that the half-chain entanglement of \(\ket{\Psi_0}\) is extensive; it scales as \(L/2\) times that of \(\ket{\psi_2}\). 
As shown in Sec.~\ref{sec3-2}, \(\ket{\Psi_0}\) admits an exact representation as a holographic isoTNS with \(\chi\geq4\)\red{. B}y contrast, it is neither \red{an} FGS\red{, a Clifford state,} nor \red{is} efficiently representable as an MPS with small bond dimension\red{, suggesting that classical simulation is intractable}. 
\red{We then evolve this state under} the critical transverse-field Ising model (TFIM):
\begin{equation}
    \hat{\mathcal{H}} = -J\sum_i\hat{\sigma}_i^z\hat{\sigma}_{i+1}^z -g\sum_i\hat{\sigma}_i^x,\quad (J,g) = (1,1).\label{eq:TFIM}
\end{equation}

\par

The results of the variational simulation are summarized in Fig.~\ref{fig4}b. 
Because the initial state exhibits volume-law entanglement, an MPS cannot provide a faithful representation \red{from} \(t=0\) \red{onward} without an exponentially large bond dimension. 
In contrast, the holographic isoTNS \red{successfully represents the short-time evolved states with a remarkably small variational error \red{at} a modest bond dimension \(\chi\).} 
\red{As the system evolves, the variational error grows steadily despite the entanglement of the target state remaining largely unchanged.} 
\red{This confirms} that the \red{bottleneck in representing} time evolution is the \red{increase} of state complexity rather than \red{the growth of} entanglement.

\red{Taken together, the holographic isoTNS is well-suited for states that are highly entangled yet have low complexity.} 
\red{Specifically, we have demonstrated its ability to represent MPS, FGS, Clifford states, extended rainbow states, and short-time-evolved states.} 
\red{While some of these admit efficient classical simulation, such methods are typically highly specialized to individual families\red{.}} 
\red{Furthermore, efficiently representing the remaining states is entirely intractable by conventional approaches.} 
\red{Thus,} \red{both the remarkable representability and the broad applicability of our ansatz stand out as key advantages.}

\section{Scalable quantum simulations}\label{sec4}

\red{Building on these} analytical and variational \red{foundations, we now shift our focus} to scalable algorithms for quantum simulation \red{using} holographic isoTNS. 

\subsection{TEBD algorithm}~\label{sec4-1}

\begin{figure}[t]
    \centering
    \includegraphics[width=\hsize]{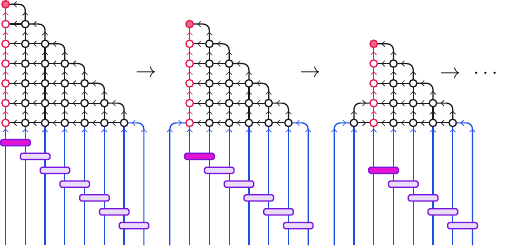}
    \caption{
    The TEBD algorithm on holographic isoTNS. 
    The time-evolution operator is approximated via the 1st-order Suzuki-Trotter decomposition into nearest-neighbor gates. 
    Gates are applied in a left-to-right sweep at odd time steps; for even time steps, a right-to-left sweep is performed (only odd time steps \red{are} shown). 
    At each gate application, the orthogonality surface is shifted accordingly to one of the sites where the gate is applied.
    }
  \label{fig5}
\end{figure}

We \red{implement} the TEBD algorithm for real-time evolution \red{on holographic isoTNS}~\cite{Vidal:PRL:2004}. 
The overall workflow is shown in Fig.~\ref{fig5}: nearest-neighbor two-qubit gates are applied sequentially, while the orthogonality surface is continually shifted to one of the two sites on which each gate acts. 
\red{Since} this shift is introduced purely for computational convenience \red{(}allowing full exploitation of the isometric constraints \red{for efficient} gate application\red{),} \red{it should leave the many-body state unchanged.} 
\red{To achieve this,} given \red{a} state \(\ket{\Psi_{\mathrm{before}}}\) with the orthogonality surface at \(i\), we \red{determine} the post-shift state \red{by solving} the variational problem \(\ket{\Psi_{\mathrm{after}}}=\mathrm{argmin}_{\ket{\Psi^\prime}}||\ket{\Psi_{\mathrm{before}}}-\ket{\Psi^\prime}||^2\).
\red{Here,} \(\ket{\Psi^\prime}\) is constrained to be a holographic isoTNS with its orthogonality surface at \(i+1\). 

In principle, attaining the optimum requires optimization over all tensors, but the associated cost is exponential. 
Instead, we optimize only the tensors in columns \(i\) and \(i+1\), keeping all others fixed. 
This is a reasonable approach because the shift alters the network structure only at these columns \red{(}see Fig.~\ref{fig2}a\red{)}. 
By virtue of the isometric constraints \red{on the unchanged} tensors, the variational problem reduces to minimizing the distance between the two affected columns, \(T^{[i,i+1]}_{\mathrm{after}}=\mathrm{argmin}_{T'}|| T_{\mathrm{before}}^{[i,i+1]} - T'||^{2}\), where \(T^{[i,i+1]}_{\mathrm{before~(after)}}\) denotes the tensors in columns \(i\) and \(i+1\) of \(\ket{\Psi_{\mathrm{before~(after)}}}\). 
The approximate shift of the orthogonality surface is diagrammatically represented as 
\begin{equation}
    \vcenter{\hbox{\includegraphics[scale=1]{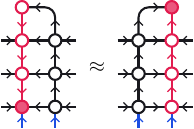}}}. \label{MM_column}
\end{equation}
Notably, this procedure can be carried out systematically using the ``Moses move" algorithm~\cite{Zaletel:PRL:2020, Sheng-Hsuan:PRB:2022} (see Appendix \ref{app:MM} for implementation details). 
In practice, we use the Moses move output as an initial guess for a subsequent alternating linearized optimization \red{scheme} that further refines the solution~\cite{Evenbly:PRB:2009,Evenbly:JStatPhys:2014,Sheng-Hsuan:PRB:2022}\red{; see Appendix~\ref{app:riemannian_optimization}. }
It is important to note that, since shifting the orthogonality surface generally introduces some error, it is crucial to minimize the number of necessary shifts in any algorithm on holographic isoTNS. 

For a given Hamiltonian, the TEBD algorithm advances the state by applying the time evolution operator \(U(\Delta t)=e^{-i\mathcal{H}\Delta t}\) over a small time step \(\Delta t\). 
A first-order Suzuki-Trotter decomposition approximates \(U(\Delta t)\) as a product of nearest-neighbor two-site gates, 
\begin{equation}
    U(\Delta t) = \prod_{i=1}^{N-1}U^{{[i,i+1]}}(\Delta t)+O(\Delta t^2),
\end{equation}
where \(U^{{[i,i+1]}}(\Delta t)\) acts locally on sites \(i\) and \(i+1\). 
Figure \ref{fig5} illustrates the gate-application sequence for an odd time step, in which the gates are applied successively from the leftmost bond to the rightmost; for an even time step, they are applied in the reverse order, from the right to the left. 
This alternating ordering minimizes the number of required shifts of the orthogonality surface within each time step. 
It is worth noting that decreasing \(\Delta t\) or using a higher-order decomposition reduces the Trotter error, but it requires additional shifts of the orthogonality surface, which increases the cumulative error in the full simulation. 

\red{Similar to the shift of orthogonality surface, t}he gate application is formulated as a variational problem. 
Given a holographic isoTNS \(|\Psi\rangle\) and a unitary gate \(U^{{[i,i+1]}}\), we search for an updated holographic isoTNS of identical structure \(|\Phi\rangle\) that best approximates the evolved state: \(\mathrm{argmin}_{\ket{\Phi}}|||\Phi\rangle-U^{{[i,i+1]}}|\Psi\rangle||^2\). 
Because \(U^{{[i,i+1]}}\) acts only on sites \(i\) and \(i+1\), we assume that tensors outside the \(i\)- and (\(i+1\))-th columns remain unchanged. 
Under the isometric constraints imposed on these uninvolved tensors, the optimization problem reduces to \(\mathrm{argmin}_{T_{\Phi}^{[i,i+1]}} || T_{\Phi}^{[i,i+1]} - U^{{[i,i+1]}}T_{\Psi}^{{[i,i+1]}} ||^{2}\), where \(T_{\Phi~(\Psi)}^{{[i,i+1]}}\) denotes the tensors on the \(i\)-th and (\(i+1\))-th columns of \(\ket{\Phi}\) (\(\ket{\Psi}\)). 
Since \(U(\Delta t)\) is close to the identity for small \(\Delta t\), 
the pre-gate tensors \(T_{\Psi}^{{[i,i+1]}}\) provide a good initial guess for the post-gate tensors \(T_{\Phi}^{{[i,i+1]}}\), which \red{are} then refined through an iterative alternating optimization \red{(}Appendix \ref{app:riemannian_optimization}\red{)}. 
Repeating this procedure for every gate at each time step following Fig.~\ref{fig5} produces the full TEBD evolution of holographic isoTNS. 

\subsection{Real-time evolution}~\label{sec4-2}

\begin{figure}[t]
    \centering
    \includegraphics[width=\hsize]{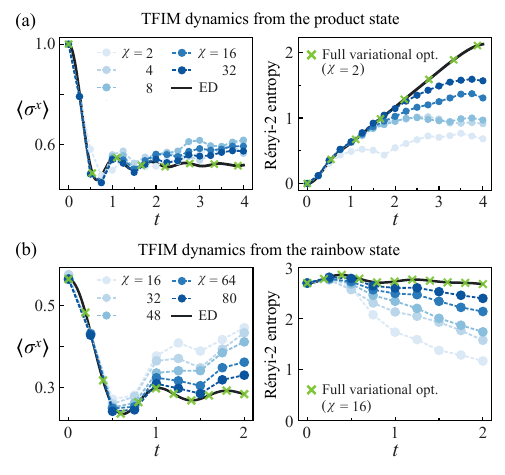}
    \caption{
        Time evolution of the mean value of \(\langle\sigma_i^x\rangle\) (left) and the R\'enyi-2-entropy \(S^{[2]}\) (right) using the TEBD algorithm. 
        The evolution is calculated under the TFIM Hamiltonian at \((J,g)=(1,1)\) in Eq.~\eqref{eq:TFIM} with the system size \(L=14\). 
        The initial state\red{s are} prepared in \red{(a)} a product state \(\ket{++\cdots ++ }\) \red{and (b)} the volume-law rainbow state in Eq.~\eqref{twoqubit}. 
        In both panels, the blue lines show the evolution of holographic isoTNS using TEBD with varying bond dimension \(\chi\), while the black curve\red{s} indicate the ED results. 
        For comparison, the results from full variational optimization with \(\chi=2\) in (a) and \(\chi=16\) in (b) are shown as green markers, which overlap with the black curves.
        The time discretization for a single TEBD step is set as \(\Delta t = 0.25\). 
    }
  \label{fig6}
\end{figure}

We benchmark the performance of the TEBD algorithm by simulating real-time evolution under the critical TFIM Hamiltonian [Eq.~\eqref{eq:TFIM}]. 
\red{We use the bond dimension $\chi$ as a proxy for both the number of parameters\red{,} scaling as $\mathcal{O}(d^4L^2 + \chi^2 d^2 L)$\red{,} and the computational cost, scaling as $\mathcal{O}(\chi^3 d^4 L^2)$ for a full TEBD update.} 
\red{To evaluate the algorithm rather than the ansatz, we restrict our analysis to cases where the representational power of holographic isoTNS is already established.} 

\red{Specifically, w}e consider two distinct initial states: (i) the fully polarized product state \(\ket{++\dots +}\) and (ii) the volume-law entangled rainbow state of Eq.~\eqref{twoqubit} with the same parameter as in Fig.~\ref{fig4}b.  
In case (i), the initial state is \red{an} FGS, and because the Hamiltonian is quadratic in its fermionic representation~\footnote{We apply the Jordan-Wigner transformation by taking the fermionic vacuum to be $\ket{++\cdots+}$; equivalently \(n_j=(1-\sigma_j^x)/2\). This choice is related to the conventional one by a global rotation and leaves the physics unchanged, while rendering Eq.~(\ref{eq:TFIM}) purely quadratic in fermions.}, the evolved states remain FGS~\cite{Surace:SciPostPhysLectNotes:2022}. 
Hence, the full time evolution can be captured already with \(\chi=2\) \red{as} discussed in Sec.~\ref{sec3-2}. 
In case (ii), we have shown variationally that holographic isoTNS reproduces short-time behavior accurately, whereas an MPS fails already at \(t=0\) (Fig.~\ref{fig4}b). 

Figure~\ref{fig6} shows the TEBD results for the spatially averaged magnetization \(\langle\sigma^x(t)\rangle=\sum_{i=1}^{L} \langle\sigma_i^x(t)\rangle /L\) and the R\'enyi-2 entropy \(S^{[2]}(t)\)\red{, comparing} them against ED results. 
We also show the results from variational optimization in green markers\red{, which we verify closely follow the ED results}. 

\red{We first focus on the case (i) in Fig.~\ref{fig6}a.} 
\red{The simulation successfully captures the short-time dynamics, confirming that our TEBD implementation performs as expected.} 
\red{However, d}espite the known representability for FGS time evolution, \red{the TEBD results for} both \(\langle\sigma^x(t)\rangle\) and \(S^{[2]}(t)\) \red{remain accurate} only up to \(t<2\)\red{, even} for our largest \(\chi=32\). 
\red{B}eyond this \red{point}, the simulation \red{diverges from} the exact dynamics, with discrepancies increasing for smaller \(\chi\). 
For reference, we confirm that \red{TEBD with} an MPS with \(\chi=32\) faithfully reproduces the dynamics over the time window considered. 

\red{Similarly, Fig.~}\ref{fig6}b displays the TEBD dynamics initialized from the highly entangled rainbow state \red{[case (ii)]}. 
\red{As in the FGS case, t}he \red{holographic isoTNS} faithfully \red{simulates} the short-time regime (e.g., for \(t\leq 0.75\) with \(\chi=80\)). 
\red{This is a notable advantage \red{because, unlike in the case (i),} an MPS at a fixed bond dimension cannot simulate \red{such volume-law entangled} evolution even for short times.}
\red{At later times}, \red{however,} deviations from the ED result emerge \red{similar to Fig.~\ref{fig6}a, and} \(S^{[2]}\) decreases from its initially large value even at high bond dimensions. 

\red{Collectively, these results demonstrate that the TEBD algorithm fails before reaching the representational limit of the ansatz.} 
\red{We attribute this} \red{premature failure} \red{to the repeated shifting of the orthogonality surface.} 
\red{This process relies on two assumptions: first, that the representability for physically relevant states is independent of the orthogonality surface position; and second, that local tensor updates around the orthogonality surface are sufficient to accurately capture the shift.} 
\red{If either assumption is violated, repeated shifts can gradually drive the state away from the exact time evolution, \red{regardless of} the representational \red{capacity} of the ansatz.} 
\red{Although variational errors in gate applications are another potential error source, they can in principle be reduced by taking a smaller \(\Delta t\).} 
\red{However, this increases the number of orthogonality surface shifts, and hence the accumulation of shift-induced errors.} 
\red{Consequently, the premature deviations expose an algorithmic limitation in the current TEBD scheme for holographic isoTNS, arising primarily from repeated shifts of the orthogonality surface.} 
Designing a scalable algorithm that fully leverages the substantial variational power of \red{the ansatz} therefore remains a significant open challenge.

\section{Conclusions} \label{sec5}

In conclusion, we introduced holographic isoTNS to represent highly entangled 1D many-body quantum states. 
Our ansatz is a (1+1)-dimensional network in which one axis represents real space while the other axis substantially enhances expressive power. 
All tensors are isometrized toward the orthogonality surface and, within that surface, toward the orthogonality center. 
Therefore, contractions involving the orthogonality \red{surface} remain efficient even in the enlarged network\red{,} consistent with the philosophy of isoTNS. 
\red{Crucially, this holographic architecture significantly expands state representability.} 
\red{Unlike standard 1D tensor networks, the} holographic isoTNS realizes volume-law entangled states \red{at a fixed bond dimension}, evidencing expressivity beyond MPS\red{,} TTN\red{, and MERA}. 
\red{We confirmed this by first showing that randomly initialized holographic isoTNS typically exhibit volume-law entanglement.}
\red{Moreover, we demonstrated through analytic arguments and variational studies that} \red{a remarkably diverse set of states}, such as MPS, FGS, Clifford states, \red{extended rainbow-states, and certain short-time evolved states} are naturally captured \red{by} our \red{single} ansatz. 
\red{The} variational analysis \red{also} reveals that increasing complexity, rather than entanglement, is the principal limiting factor for \red{the} \red{state representability}. 
\red{T}herefore\red{, we} conclude that holographic isoTNS is particularly well suited to highly entangled, low-complexity states\red{.}
\red{Because} \red{this} regime \red{covers} many physically relevant situations \red{that traditionally defy efficient simulation}\red{,} \red{our ansatz promises to have a} broad impact \red{on} quantum many-body studies \red{especially in the volume-law regime}. 

To exploit this capacity, we further implemented a TEBD scheme tailored to the ansatz with a computational cost scaling quadratically in system size. 
However, we found that errors accumulated over TEBD sweeps quickly lead to deviations from exact dynamics. 
Designing scalable algorithms that preserve this efficiency while controlling error growth remains a compelling direction for future work. 
\red{One promising strategy is the optimization of the holographic isoTNS via variational Monte Carlo \red{(VMC)} methods, which have recently demonstrated success in the optimization and time evolution of PEPS~\cite{wu_algorithms_2025, chen_variational_2026, wu_real-time_2026}.}
\red{Notably, the isometric form of the holographic isoTNS enables a more efficient direct sampling of the probability distribution corresponding to the represented quantum state~\cite{dektor_sampling_2026}\red{, making VMC a natural avenue for unlocking the full potential of our ansatz}.}

Several generalizations of holographic isoTNS can be made. 
\red{The most straightforward is the extension to higher spatial dimensions.} 
For \red{example}, a (2+1)-dimensional holographic isoTNS on a cubic lattice would naturally represent volume-law entangled states in 2D many-body systems~\cite{Tepaske:PRR:2021}. 
Infinite systems with translational invariance offer another compelling arena. 
\red{Such} states \red{could be constructed} by applying a finite number of \red{unitary tensor} layers to an infinite MPS with a given unit cell\red{~\cite{Wu:PRB:2023}.} 
Extensions to fermionic degrees of freedom are likewise promising~\cite{Dai:PRL:2025, Yantao:PRXQ:2025}. 
In 1D, the Jordan-Wigner transformation maps fermions to qubits\red{.} 
\red{While this may} introduce nonlocal strings, \red{they} do not pose a major obstacle for our ansatz \red{due to its inherent ability to} capture large entanglement. 
In higher dimensions, faithfully encoding fermionic anticommutation within the holographic structure \red{remains} nontrivial. 
Overall, we emphasize that the existing body of work on isoTNS provides strong guidance for these extensions of holographic isoTNS.
\par
The quantum circuit viewpoint suggests a complementary route to extending holographic isoTNS~\cite{Slattery:arXiv:2021}. 
Because the isometric constraints impose a directed flow of information, the network admits a natural mapping to a quantum circuit. 
\red{As demonstrated by numerous recent works, s}uch quantum circuit ansätze are known to be highly expressive despite their sparsity~\cite{Sheng-Hsuan:PRXQ:2021,Haghshenas:PRX:2022,Sajant:PRXQ:2023}. 
This perspective motivates an inverse construction: Using a given quantum circuit ansatz as a blueprint to define an alternative holographic isoTNS structure that encodes it.
In particular, circuits with nonlocal gates, multi-qudit gates, or mid-circuit measurements could be embedded naturally within an extended holographic isoTNS. 
The notion of an orthogonality surface can be modified more flexibly, potentially enabling error-free orthogonality surface shifts. 
Systematically exploring holographic isoTNS configurations in tandem with quantum circuit ansätze is therefore a promising direction for future work. 

\section*{Acknowledgments}
The authors thank Sheng-Hsuan Lin, Sajant Anand, Ra\'ul Morral-Yepes, Marc Langer, Bernhard Jobst, Masataka Kawano, Johannes Hauschild, and Yukitoshi Motome for fruitful discussions. We particularly thank Michael P Zaletel for discussions and related collaborations. 
K.K. was supported by the Program for Leading Graduate Schools (MERIT-WINGS) and JST BOOST (No.~JPMJBS2418). 
This work was supported by the European Research Council (ERC) under the European Union's Horizon 2020 research and innovation program under Grant Agreements No.~771537 and No.~851161, the Deutsche Forschungsgemeinschaft (DFG, German Research Foundation) under Germany's Excellence Strategy EXC-2111-390814868, TRR360-492547816, and the Munich Quantum Valley, which is supported by the Bavarian state government with funds from the Hightech Agenda Bayern Plus.

\textbf{Data and materials availability}. 
Data analysis and simulation codes are available on Zenodo upon reasonable request~\cite{kobayashi_2025_17907383}.

\appendix

\section{Variational optimization of holographic isoTNS}
\label{app:riemannian_optimization}
In the following we show how one can optimize a holographic isoTNS to minimize a given cost function. We start by defining the problem as a general optimization problem, and proceed to give details on two specific methods to solve the problem.  
Consider the cost function 
\begin{equation*}
    f:\mathbb{C}^{n_1\times m_1}\times\mathbb{C}^{n_2\times m_2}\times\cdots\times\mathbb{C}^{n_N\times m_N} \times \mathbb{C}^{p} \to \mathbb{R},
\end{equation*}
mapping $N+1$ tensors to a scalar cost value. This cost function can, for instance, represent the contraction of a tensor network consisting of $N$ isometric tensors $W_i\in\mathbb{C}^{n_i\times m_i}$ with $n_i\ge m_i$ and (optionally) an orthogonality center tensor $T\in\mathbb{C}^p$. \red{Here,} we have reshaped the isometric tensors into isometric matrices and the orthogonality center into a vector. 
Examples for such cost functions $f$ are the overlap of a holographic isoTNS with a reference wavefunction \red{$\braket{\Psi_\text{ref}|\Psi_\text{isoTNS}}$} \red{(to be maximized)} or the energy $\bra{\Psi_\text{isoTNS}}\hat{\mathcal{H}}\ket{\Psi_\text{isoTNS}}$ of a holographic isoTNS for a model Hamiltonian $\hat{\mathcal{H}}$ \red{(to be minimized)}. 
\red{In the following, we focus on the maximization case for clarity.} 
We are interested in the constrained optimization problem
\begin{equation*}
    W_1^\text{opt}, W_2^\text{opt}, \dots, W_N^\text{opt}, T^\text{opt} = \underset{W_1, W_2, \dots, W_N, T}{\argmax} f(W_1, W_2, \dots, W_N, T)
\end{equation*}
under the constraints $W_i^\dagger W_i = \id$ for isometries and $\lVert T\rVert = 1$ for the orthogonality center. In the following we discuss two methods for solving this optimization problem, namely an alternating linearized optimization scheme due to Evenbly and Vidal~\cite{Evenbly:PRB:2009,Evenbly:JStatPhys:2014}, and Riemannian optimization of product manifolds.

\subsection{Alternating linearized optimization}

Assume first that the cost function is linear in all its arguments, i.e., given by the contraction of a tensor network in which each tensor appears only once. In this case, one can extremize the cost function by iteratively optimizing one tensor at a time. To optimize an isometry $W_i\in\mathbb{C}^{n_i\times m_i}$, we first contract the tensor network without $W_i$ into an environment tensor $E\in\mathbb{C}^{n_i\times m_i}$. The optimal tensor $W_i^\prime$ when keeping all other tensors fixed is then given by
\begin{equation}
    W_i^\prime = \underset{W_i^\dagger W_i = \id}{\argmax}\Tr(W_i^\dagger E).
\end{equation}
This optimization problem is known as the \textit{orthogonal Procrustes problem} and permits the closed form solution
\begin{equation}
    W_i^\prime = UV^\dagger,
    \label{eq:orthogonal_procrustes_solutioni_isometry}
\end{equation}
where the matrices $U$ and $V$ are defined through the SVD $E = USV^\dagger$. Similarly, if the tensor to be optimized is the orthogonality center $T$, contracting all other tensors into an environment $E\in\mathbb{C}^p$ yields the optimization problem
\begin{equation}
    T^\prime = \underset{\lVert T\rVert = 1}{\argmax}\Tr(T^\dagger E),
\end{equation}
which is solved by
\begin{equation}
    T^\prime = E/\lVert E\rVert.
    \label{eq:orthogonal_procrustes_solutioni_sphere}
\end{equation} 
For derivations of the closed form solutions, see the appendices of Refs.~\cite{Benjamin:arXiv:2025, Sheng-Hsuan:PRB:2022}. \par
Each of these local updates is optimal and thus the value of the cost function is never decreased. Therefore, \red{i}f the cost function is linear in all tensors and bounded, $f(W_1,\dots,W_N,T) \le c\in\mathbb{R}$, the algorithm is guaranteed to converge to a local maximum.\par
In general the cost function $f$ is not linear. Non-linear cost functions are for example encountered when optimizing a MERA wavefunction or during the disentangling step in the Moses move (see Appendix \ref{app:MM}). 
Assume that the tensor $W_i$ appears more than once in the tensor network describing the cost function. It was proposed by Evenbly and Vidal~\cite{Evenbly:PRB:2009,Evenbly:JStatPhys:2014} in the context of MERA to linearize the cost function and to update the tensor $W_i$ iteratively such that the linearized cost function is \red{maximized} in each iteration. For this, we contract all tensors except one of the tensors $W_i$ into an environment $E(W_i)\in\mathbb{C}^{n_i\times m_i}$, which now depends on $W_i$. We then update $W_i$ by treating the environment as if it were independent of $W_i$ and employing the closed form update Eq.~\eqref{eq:orthogonal_procrustes_solutioni_isometry}. This is repeated until $W_i$ is converged. However, if and how fast $W_i$ converges depends on the cost function, and convergence cannot be guaranteed in general.

\subsection{Riemannian Optimization}

The alternating linearized optimization can only be used if the cost functions can be expressed as the contraction of a tensor network. 
Additionally, the method often converges only very slowly, especially if the cost function depends on many tensors. 
\red{As an alternative, we consider} Riemannian optimization\red{, which} generalizes known \red{E}uclidean optimization algorithms like Gradient Descent, Conjugate Gradients (CG) and quasi-newton methods like the Trust Region Method (TRM) to \textit{Riemannian manifolds}. 
In the following, we give a brief introduction to Riemannian optimization. For an in-depth introduction to the topic we recommend Refs.~\cite{absil_optimization_2008, luchnikov_riemannian_2021, hauru_riemannian_2021}. \par
Let us first assume that the cost function depends only on a single tensor, which can either be constrained to be an isometry or a tensor of norm one. 
The space of all $n\times m$ isometries $W$ is a Riemannian manifold called the \textit{Stiefel manifold} $\Stiefel$. 
Likewise, the complex unit sphere $S^{2p-1} \subset \mathbb{C}^p$, i.e., the space of all vectors $T\in\mathbb{C}^p$ of norm one, $\lVert T\rVert = 1$, is a Riemannian manifold. 
In the following, let $\mathcal{M}$ denote the manifold which we want to optimize over. 
Standard optimization algorithms iteratively improve an iterate $X_k\in\mathcal{M}$, $k = 1, 2,\dots$ until a local extremum of the cost function $f : \mathcal{M}\to\mathbb{R}$ is reached. 
At each step, the iterate is moved along an update direction $\xi_k$ with a step size $\alpha_k$. 
The update direction and step size are typically computed from the current and past gradients $\nabla f(X_k)$ and sometimes higher derivatives of the cost function. 
In optimization algorithms defined on Euclidean vector spaces, the update direction lies in the same vector space as the iterates $X_k$, and the iterate can simply be \red{updated} as $X_{k+1} = X_k + \alpha_k\cdot\xi_k$. 
When optimizing over a Riemannian manifold, the update direction should first be projected into the tangent space $T_{X_k}\mathcal{M}$ to the manifold $\mathcal{M}$ at point $X_k\in\mathcal{M}$ \cite{absil_optimization_2008}. 
Further, just updating the iterate by moving the current iterate along the update direction would in general leave the manifold. To ensure that we stay on the manifold, $X_{k+1}\in\mathcal{M}$, one can introduce a \textit{retraction} $R_{\xi_k}:\mathbb{R}\to\mathcal{M}$. 
One can think of $R_{X_k}(\alpha_k)$ as moving along the direction of $\xi_k$, starting from $X_k$, with a step size of $\alpha_k$, while staying on the manifold $\mathcal{M}$. 
A practical retraction for the Stiefel manifold is given by ${R_{\xi_k}(\alpha_k)=\qf(X_k+\alpha_k\xi_k)}$~\cite{absil_optimization_2008}, where $\qf(A)$ denotes the $Q$-factor of the QR-decomposition $A = QR$.
For the complex unit sphere, the \red{retraction} ${\red{R_{\xi_k}(\alpha_k)} = (X_k+\alpha_k\xi_k)/\lVert X_k+\alpha_k\xi_k\rVert}$ can be used~\cite{absil_optimization_2008}.\par
Some algorithms, e.g. CG, need gradients of the cost functions at previous iterates for computing the update direction at the current iterate. 
Since the tangent spaces at two iterates $X_k$ and $X_{k^\prime}$ are generally different, previous gradients need to first be brought to the tangent space of the current iterate using a \textit{vector transport} $T_{k\to k^\prime}:T_{X_k}\mathcal{M}\to T_{X_{k^\prime}}\mathcal{M}$~\cite{absil_optimization_2008}. 
A particularly simple vector transport for the Stiefel manifold and the complex unit sphere can be implemented by simply projecting gradients at previous iterates to the tangent space of the current iterate. 
Lastly, second order algorithms such as the TRM require the generalization of the Hessian-vector product to Riemannian manifolds, which is given by the \textit{Riemannian connection}. 
For the Stiefel manifold (the complex sphere), the Hessian-vector product can be computed by projecting the Hessian-vector product from the embedding vector space $\mathbb{C}^{n\times m}$ ($\mathbb{C}^p$) to the manifold.\par
In this work, we used two optimization algorithms, CG and the TRM. CG uses accumulated gradients from previous iterates to compute improved orthogonal search directions. 
The TRM locally approximates the cost function using a quadratic model within a neighborhood of radius $\Delta_k$ around the current iterate $X_k$.
This approximate model is then minimized inside the trust region using truncated Conjugate Gradients (tCG), which is particularly efficient for quadratic problems. 
Based on how well the model approximates the true cost function at the current iterate, the trust region can be either expanded or contracted. 
The TRM combines the advantages of local superlinear convergence with certain global convergence guarantees \cite{absil_optimization_2008}, and can be viewed as an enhanced version of Newton's method.
For a more in-depth discussion of CG and the TRM, see Refs.~\cite{absil_optimization_2008, zhu_riemannian_2017, hauru_riemannian_2021, absil_trust-region_2007, townsend_pymanopt_2016}.
In this work, we use the CG and TRM algorithms implemented in the \textbf{pymanopt} package~\cite{townsend_pymanopt_2016}. Gradients and higher order derivatives are computed using the \textbf{autograd} feature in \textbf{jax}~\cite{jax2018github}.

\section{Moses move}\label{app:MM}

\begin{figure}[t]
    \centering
    \includegraphics[width=\hsize]{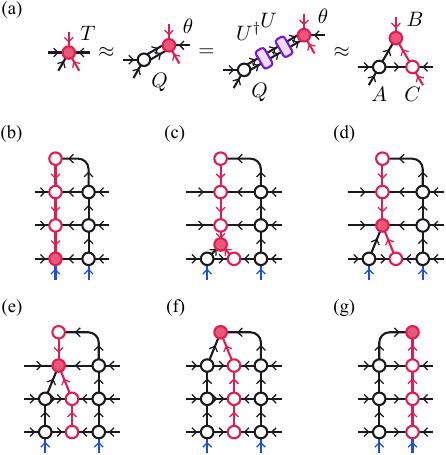}
    \caption{
        A sketch of the Moses move algorithm. 
        (a) The tripartite decomposition, as defined in Eq.~\eqref{eq:tripartite_decomposition}. 
        (b)-(g) Steps for shifting the orthogonality surface one column to the right. 
        (b) Initially, the orthogonality surface is on the left column. 
        (c)-(f) The tripartite decomposition is repeatedly applied to unzip the column. 
        (g) The intermediate column is contracted into the right column, completing the shift of the orthogonality surface. 
    }
    \label{fig7}
\end{figure}

As emphasized in the main text, the strength of holographic isoTNS stems from efficient contraction outside the orthogonality surface\red{, enabled by} the isometric conditions. 
Hence, shifting the orthogonality surface is of fundamental importance for our ansatz. 
The Moses move provides a systematic prescription for this shift~\cite{Zaletel:PRL:2020, Sheng-Hsuan:PRB:2022}. 
The core step of the Moses move is a tripartite tensor decomposition of the orthogonality center tensor, sketched in Fig.~\ref{fig7}a. 
We denote the tensor to be decomposed by \(T_{l, b, a, r, \alpha}\) with subscripts labeling the left, below, above, and right legs, and the down-right ancillary leg, respectively. 
A detailed account of the meaning of these legs is postponed until after the full procedure has been outlined\red{.} 
\red{F}or the moment we assume that the left, right, and below legs have bond dimension \(1\) or \(d\), while those of the above and ancillary legs are bounded by \(\chi\).

The tripartite decomposition begins with a QR decomposition of the bipartition \(T_{(l, b), (a, r, \alpha)} = Q_{(l, b), \beta}\theta_{\beta,(a, r, \alpha)}\). 
Here \(Q\) is an isometry, and the outgoing leg \(\beta\) has the smaller dimension of the two grouped subspaces \((l, b)\) and \((a, r, \alpha)\); hence \(\mathrm{dim}~\beta\) is \(1\), \(d\) or \(d^2\) under the assumption introduced above. 
We focus on the latter two cases and further split \(\beta\) into two indices, \(\beta_u\) and \(\beta_r\): when \(\mathrm{dim}~\beta = d^2\) we set \(\mathrm{dim}~\beta_u = d\) and \(\mathrm{dim}~\beta_r = d\), whereas for \(\mathrm{dim}~\beta = d\) we take \(\mathrm{dim}~\beta_u = 1\) and \(\mathrm{dim}~\beta_r = d\). 
Importantly, at this stage a unitary \(U\) can be inserted as a gauge degree of freedom: \(T=(QU^\dagger)(U\theta)\equiv A\theta'\). 
A suitable choice of the disentangling unitary \(U\) redistributes the singular values and can substantially reduce the truncation error introduced in the next step. 
We then apply an SVD to the disentangled tensor \(\theta'_{(\beta_u, a), (\beta_r, r, \alpha)} \simeq B_{(\beta_u, a), \gamma}C_{\gamma, (\beta_r, r, \alpha)}\), where \(C\) is an isometry and the bond dimension of \(\gamma\) is truncated to the largest \(\chi\) singular values. 
Collecting these factors yields the tripartite decomposition shown in Fig.~\ref{fig7}a, 
\begin{equation}
    T_{l, b, a, r, \alpha}\simeq A_{l, b, \beta_a,\beta_r} B_{\beta_a, a, \gamma} C_{\gamma,\beta_r, r, \alpha}. \label{eq:tripartite_decomposition}
\end{equation}
The accuracy of the decomposition depends sensitively on the disentangling unitary. 
Following Ref.~\cite{Sheng-Hsuan:PRB:2022}, we select the unitary that minimizes the R\'enyi-\(1/2\) entropy of the bipartition of \(\theta'\) using a Riemannian-manifold optimization method. 
Because R\'enyi-\(n\) entropy with \(n<1\) bounds the truncation error at a fixed bond dimension, this criterion provides an effective means of reducing the loss of accuracy~\cite{Verstraete:PRB:2006}.

We are now ready to detail the Moses move protocol. 
For later reference, let \(\{\Psi^{[h]}\}\) denote the tensors on the orthogonality surface column and \(\{\Phi^{[h]}\}\) those in the neighboring right column, where the position label \(h\) is counted upward from the bottom. 
The first step shifts the orthogonality center to the bottom tensor \(\Psi^{[h=1]}\); this is an exact gauge transformation that introduces no error (Fig.~\ref{fig7}b). 
We then rewrite \(\Psi^{[h=1]}\) as a five-leg tensor \(T_{l,b,a,r,\alpha}\): the three incoming virtual legs from the left, right, and above become the indices \(l\), \(r\), and \(a\), the physical leg is relabeled \(b\), and a trivial auxiliary leg \(\alpha\) is appended. 
By construction, the bond dimensions of \(l\), \(r\), and \(b\) legs are either \(1\) or \(d\), while those of \(a\) and \(\alpha\) are upper bounded by \(\chi\), consistent with the assumption introduced above. 
Next, we apply the tripartite decomposition of Eq.~\eqref{eq:tripartite_decomposition} to \(T\), obtaining three tensors \(A^{[h=1]}, B^{[h=1]}\), and \(C^{[h=1]}\) (Fig.~\ref{fig7}c). 
Among the resulting legs, only above (\(a\)) leg and the newly introduced leg (\(\gamma\)) have the bond dimension \(\chi\); all others retain bond dimensions \(1\) or \(d\). 
The tensor \(B^{[h=1]}\) is then contracted with the tensor directly above \(\Psi^{[h=2]}\) along the \(a\) leg. 
This produces a new five-leg tensor one layer higher, whose legs all point inward and respect the assumed bond dimensions (Fig.~\ref{fig7}d). 
Iterating this sequence raises the orthogonality center layer by layer until it reaches the topmost tensor (Figs.~\ref{fig7}e and \ref{fig7}f). 
At this point, the orthogonality surface column has been split into two columns, \(\{A^{[h]}\}\) on the left and \(\{C^{[h]}\}\) on the right. 
Finally, each \(C^{[h]}\) is contracted with the adjacent tensor \(\Phi^{[h]}\); truncating the resulting vertical bond to dimension \(\chi\) completes the shift of the orthogonality surface by one column (Fig.~\ref{fig7}g). 
We note that the cost of the Moses move is on the order of \(O(\red{L}\chi^3)\). 
As discussed in Sec.~\ref{sec4-1}, we use the outcomes of the Moses move as an initial guess for the subsequent alternating linearized variational optimization, in which the solution is refined to reduce errors in the shift~\cite{Sheng-Hsuan:PRB:2022}.

\bibliography{biblio}

\begin{thebibliography}{95}
\bibitem [{com()}]{complexity}%
  \BibitemOpen
  \href@noop {} {}\bibinfo {note} {There are different ways a complexity measure on quantum states can be defined. We adapt the following definition: The complexity of a quantum state $\ket{\Psi}$ is the minimal circuit depth needed to transform the computational basis state to a state $\ket{\Psi^\prime}$ with ${\lVert\ket{\Psi}-\ket{\Psi^\prime}\rVert <\varepsilon}$ using only local two-qubit unitaries.}\BibitemShut {Stop}%
\end{thebibliography}

\end{document}